%% file: MoS2_ribbon_final.tex
\documentclass[journal=prl,manuscript=article,showpacs,twocolumn,superscriptaddress]{revtex4-1}

\usepackage{graphicx}
\usepackage{amssymb}

\usepackage{amsmath}
\usepackage{comment}
\usepackage[colorlinks=true,linkcolor=blue,citecolor=blue,urlcolor=blue]{hyperref}
\usepackage{tabularx}
\usepackage{color}
\definecolor{Blu}{rgb}{0.00,0.00,1.00}
\definecolor{Red}{rgb}{1.00,0.00,0.00}
\definecolor{Orange}{rgb}{1.00,0.40,0.00}
\definecolor{DarkGreen}{rgb}{0.00,1.00,0.00}

\def\QE{{\sc Quantum ESPRESSO}}
\def\yambo{{\sc yambo}}

\AtBeginDocument{\usepackage{booktabs}}
\begin{document}

\title{Intrinsic edge excitons in two-dimensional MoS$_2$}

\author{Pino D'Amico}
\email{pino.damico@nano.cnr.it}
\affiliation{Centro S3, CNR--Istituto Nanoscienze, Via Campi 213/A, I-41125 Modena, Italy }
\author{Marco Gibertini}
\email{marco.gibertini@epfl.ch}
\affiliation{Department of Quantum Matter Physics, Universit\'e de Gen\`eve, CH-1211 Geneva, Switzerland}
\affiliation{Theory and Simulation of Materials (THEOS) and National Centre for Computational Design and Discovery of Novel Materials (MARVEL), \'Ecole Polytechnique F\'ed\'erale de Lausanne, CH-1015 Lausanne, Switzerland}
\author{Deborah Prezzi}
\email{deborah.prezzi@nano.cnr.it}
\affiliation{Centro S3, CNR--Istituto Nanoscienze, Via Campi 213/A, I-41125 Modena, Italy }
\author{Daniele Varsano}
\affiliation{Centro S3, CNR--Istituto Nanoscienze, Via Campi 213/A, I-41125 Modena, Italy }
\author{Andrea Ferretti}
\affiliation{Centro S3, CNR--Istituto Nanoscienze, Via Campi 213/A, I-41125 Modena, Italy }
\author{Nicola Marzari}
\affiliation{Theory and Simulation of Materials (THEOS) and National Centre for Computational Design and Discovery of Novel Materials (MARVEL), \'Ecole Polytechnique F\'ed\'erale de Lausanne, CH-1015 Lausanne, Switzerland}
\author{Elisa Molinari}
\affiliation{Dipartimento di Fisica Informatica e Matematica, Universit\`a di Modena e Reggio Emilia, Via Campi 213/a I-41125 Modena, Italy}
\affiliation{Centro S3, CNR--Istituto Nanoscienze, Via Campi 213/A, I-41125 Modena, Italy }

\begin{abstract}


Using accurate first-principles calculations based on many-body perturbation theory we predict that two-dimensional MoS$_2$ hosts edge excitons with universal character, intrinsic to the existence of edges and lying well below the onset of bulk features. These excitons are largely insensitive to edge terminations or orientation, persisting even in the presence of metallic screening at zigzag edges, with large binding energies of $\sim$0.4 eV. Additional excitons can also emerge in ultranarrow ribbons, or as a function of the chemical nature of the termination. The chemical, structural, and electronic similarities with Se- or W-based transition-metal dichalcogenides suggest that these optical features could be common in this class of materials.

\end{abstract}

\maketitle


In the field of two-dimensional (2D) materials, transition-metal dichalcogenides (TMD) have attracted great interest in the last few years~\cite{Wang2012,Manzeli2017} thanks to the wide range of physical properties they exhibit, from superconductivity~\cite{Ye2012,Xi2015b,Costanzo2016} to charge density waves~\cite{Xi2015,Yu2015}, topological~\cite{Qian2014,Fei2017,Tang2017,Wu2018} and excitonic order~\cite{varsano19}.
Among these, group-VI TMDs, based on either tungsten or molybdenum, were the first semiconducting 2D materials to be explored~\cite{Novoselov2005,Radisavljevic2011}. 
Their direct band gap~\cite{mak2010} enhances light-matter interactions, revealing a wealth of optical properties: photoluminescence~\cite{mak2010,splendiani2010}, radiative processes~\cite{palummo15, pogna16},  dark excitons~\cite{berkelbach15, baranowski17, molas17,malic18}, exciton-plasmon interactions~\cite{ tokman15, moody16,kang17, singha17, nerl17,dinh17}, chiral and defect effects \cite{gong17, yu14, dubey17, gogoi17}, exciton tuning through substrate~\cite{mertens14,jia17} and interlayer interactions~\cite{Fang2014,chen16,mouri17,latini17,Kunstmann2018}, as well as complex excitations like trions~\cite{jadczak17,  kim16, lin14, zhang14} and biexcitons~\cite{Mai2014,Shang2015,You2015}. 

Although initially most efforts were devoted to the investigation of  extended 2D TMDs, it was soon realized that in finite-size systems, such as nanoribbons or triangular islands, mid-gap states emerge and are localized at the edges~\cite{Helveg2000,Bollinger2001,Bollinger2003,Zhou2013}. As a consequence, the energy gap is significantly reduced close to the edge, but still remains finite for armchair (AC) terminations~\cite{Bollinger2003,gibertini15}. On the contrary, in zigzag (ZZ) edges the gap closes and the edge states become metallic, thus forming one-dimensional (1D) wires of free carriers~\cite{Bollinger2001,Bollinger2003,gibertini15}. This different behaviour is due to the polar nature of TMDs that gives rise to a polar discontinuity across ZZ edges; this discontinuity smoothly decreases as a function of orientation, disappearing for AC terminations~\cite{Guller2013,gibertini14,gibertini15,Guller2015}. Free carriers at the edges will emerge above a critical width to screen the  bound polarization charges associated with the polar discontinuity~\cite{MartinezGordillo2015}. This behaviour is analogous to the emergence of a 2D electron gas at the interface between bulk insulating perovskites such as LaAlO$_3$ and SrTiO$_3$~\cite{Ohtomo2004}. Remarkably, the existence of edge states in TMDs has been confirmed in experiments using scanning tunneling spectroscopy~\cite{Helveg2000,Bollinger2001}, with additional experimental evidence that can be inferred from the observation of edge-localized room-temperature photoluminescence~\cite{Gutierrez2013},  non-linear optical responses~\cite{Yin2014,Lin2018}, and  enhanced electrochemical H$_2$ evolution~\cite{Jaramillo2007}. 

Such metallic edge states have the potential to host promising applications, as they are predicted, among other things, to become magnetic~\cite{Vojvodic2009,Botello2009,pan12,nam17}, to sustain edge plasmons~\cite{Andersen2014}, to bind efficiently Li atoms~\cite{Li2012}, and to exhibit catalytic properties~\cite{Lauritsen2003,Liu2017}. This is particularly relevant for nanoribbons (NRs): edges play a dominant role in determining their properties, which are then sensitive to functionalizations~\cite{xiao15, li14}, edge terminations~\cite{pan12} and edge roughness~\cite{babaee17,ridolfi17}. A  systematic exploration of such theoretical predictions could  now be experimentally achievable thanks to recent developments in the realization of atomically flat edges~\cite{Chen2017}, including the  growth of nanoribbons, either isolated~\cite{li18} or embedded in a different 2D material~\cite{Han2017,sahoo18}.

\begin{figure}
\begin{center}
\includegraphics[width=0.48\textwidth]{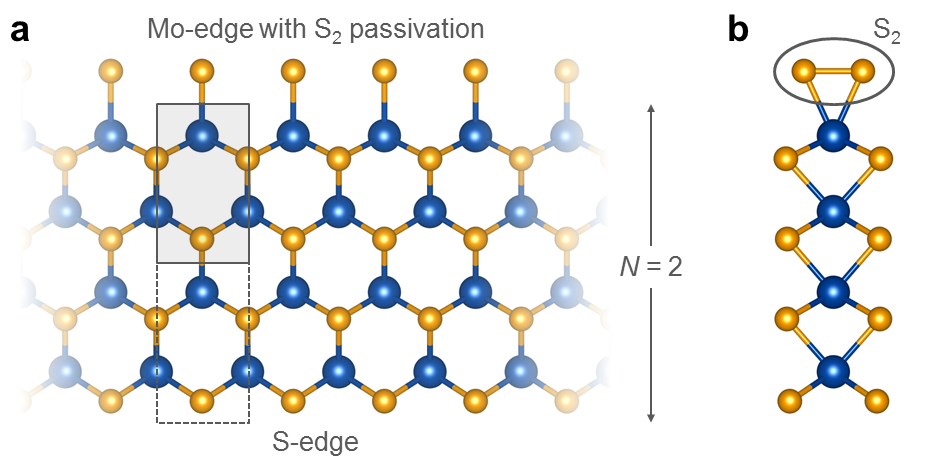}
\caption{Ball-and-stick model of the top (a) and lateral (b) view of a $N=2$ MoS$_2$ ZZ-NR, with the Mo-edge passivated with S-dimers. Mo and S are represented in blue and orange, respectively. In (a) we highlight with gray shading the non-primitive orthorhombic cell of 2D MoS$_2$: its repetition along the armchair direction defines the width parameter $N$ for zigzag NRs, i.e. $N=2$ in this case.
\label{atomic_structure}}
\end{center}
\end{figure}

In this Letter we use accurate first-principles many-body  approaches based on  many-body perturbation theory ($GW$) and the Bethe-Salpeter Equation (BSE)~\cite{onid+02rmp} to investigate the impact of edge states on the optical properties of finite-size TMDs.
By considering MoS$_2$ ZZ-NRs as a prototypical example, we argue for the presence of additional edge excitons that are clearly distinguishable from bulk spectral features of the extended MoS$_2$ monolayer.
These excitons arise from edge-localized states and can be classified as inter- and intra-edge excitons. Despite the metallic character of the edges, we predict a large binding energy in both cases, as a consequence of ineffective screening in reduced dimensionality. The analysis of the width dependence of $GW$-BSE spectra shows that inter-edge excitons disappear for sufficiently large NRs, while intra-edge features  are robust and would be accessible in experimentally available samples~\cite{wei17}.
\begin{figure}
\begin{center}
\includegraphics[width=0.48\textwidth]{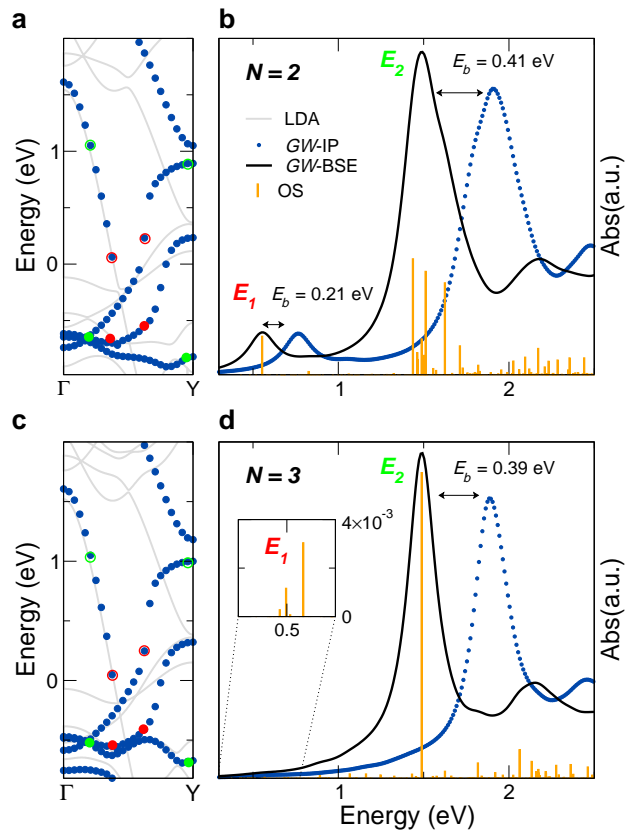}
\caption{Electronic band-structures (a,c) and optical absorption spectra (b,d) for the $N=2$ (a,b) and $N=3$ (c,d) MoS$_2$ ZZ-NRs, respectively.
Panels a and c: Gray solid lines and blue dots correspond to the LDA Kohn-Sham and to the $G_0W_0$ bands, respectively, where the Fermi level has been set to zero.
Panels b and d: Blue dotted curves ($GW$-IP) represent the optical absorption spectra as obtained according to the independent particle approximation by considering the $GW$ corrected energies and the LDA Kohn-Sham orbitals, whereas black solid curves ($GW$-BSE) correspond to the optical absorption spectra with the inclusion of electron-hole interaction within the $GW$-BSE scheme. In both cases, an artificial Lorentzian broadening of 0.1 eV has been used for convenience, while the relative oscillator strength of individual excitonic states is indicated by orange bars. The binding energies $E_b$ of the most relevant peaks, i.e. E$_1$ and E$_2$, are indicated in panels b and d, while the peak composition in terms of single-particle transitions is indicated by red (E$_1$) and green (E$_2$) circles in panels a and c, where full (empty) symbols stand for occupied (empty) states.
\label{bands_and_absorption_spectra}}
\end{center}
\end{figure}
We choose isolated MoS$_2$ NRs with ZZ edges, not only because the polar discontinuity and the metallic character are maximized, but also because these are expected to be more stable~\cite{Helveg2000,schweiger2002,Bollinger2003,Lauritsen2007} and are indeed typically observed in experiments~\cite{Helveg2000,Zhou2013,Gutierrez2013,Chen2017}. As shown in Fig.~\ref{atomic_structure}, such NRs can be considered as resulting from a finite number $N$ of repetitions of the bulk 2D orthorhombic cell (non-primitive) along the transverse AC direction, then replicated indefinitely along the longitudinal ZZ direction.  As a result of the non-centrosymmetric nature of the parent material, ZZ NRs then naturally display a S- and a Mo-terminated edge. 
In the following we will consider bare S-edges and Mo-edges passivated with sulfur dimers (see Fig.~\ref{atomic_structure}), which were predicted to be the most stable configurations in standard experimental conditions~\cite{biskov1999, schweiger2002}. 
Alternative terminations of the Mo-edge, such as bare or S-monomer passivation, are also investigated and the corresponding results are reported in the Supplemental Material (SM)~\cite{SM}. As we shall discuss below, the novel excitonic features we have identified are actually independent of the precise edge termination. 

We start by considering the band structure of narrow-width ZZ-NRs obtained by using the local-density approximation (LDA) to density-functional theory as implemented in \QE{}~\cite{espresso09,espresso17} and by including many-body corrections within the so-called $G_0W_0$ approach~\cite{onid+02rmp} using the \yambo{} code~\cite{yambo,sangalli2019many}. Figure \ref{bands_and_absorption_spectra} displays the LDA (gray lines) and $G_0W_0$ (blue dots) band structure for $N=2$ (a) and $N=3$ (c) MoS$_2$ ZZ-NRs, corresponding to widths of $1.11$ and $1.65$ nm respectively. Irrespective of the level of theory and of the width of the NR, three edge-localized states cross the Fermi level, arising from the polar discontinuity between MoS$_2$ and vacuum across the edge of the NR ~\cite{gibertini14,gibertini15}. The comparison between the results for $N=2$ and $N=3$ shows that these edge-localized bands are nearly independent of the NR width, at variance with the bulk extended states. 

The different spatial localization results into different self-energy corrections to the LDA bands when Coulomb interactions are taken into account at the $G_0W_0$ level. An accurate treatment beyond the simplified `scissor-and-stretching' correction scheme is thus required for a reliable description of the whole quasi-particle band-structure (see details in the SM~\cite{SM}). 

The optical absorption spectra for $N=2$ (b) and $N=3$ (d) MoS$_2$ ZZ-NRs are also reported in Fig.~\ref{bands_and_absorption_spectra}, beside the corresponding band structures. The blue dotted curves indicate the optical spectra obtained without the effect of electron-hole (e-h) interactions, thus directly reflecting the independent-particle optical transitions ($GW$-IP) and the energy differences of the $G_0W_0$-corrected bands depicted with blue dots in panels a and c, respectively. The black curves represent the spectra obtained by including e-h interactions within the $GW$-BSE scheme, where the effect of the metallic limit $q\rightarrow0$ on the screened Coulomb interaction has been taken into account (see SM \cite{SM} for further  details). For the latter, we also report (with orange bars) the individual excitonic states obtained from the diagonalization of the BSE Hamiltonian, where the bar height encodes the relative optical strength of the excitations. 

Focusing on the low-energy part of the spectrum, namely on the region below 2 eV, we note that for the thinner system ($N=2$) two peaks appear, hereafter labelled E$_1$ and E$_2$. They are located at 0.55 and 1.49 eV, i.e.\ well-below the optical onset of the 2D parent material which is dominated by two absorption peaks (usually labelled A and B) at 1.88 and 2.03 eV~\cite{mak2010}.
In addition, the comparison with the $GW$-IP spectrum allows us to extract the binding energy $E_b$ of each peak, which amounts to 0.21 and 0.41 eV, respectively, much smaller than that of the A and B peaks for 2D-MoS$_2$ that is close to 1 eV~\cite{Qiu2013} in vacuum. 
These observations suggest that E$_1$ and E$_2$ have a rather different origin with respect to the A and B peaks of 2D-MoS$_2$. This is further supported by the fact that bulk-like excitons, with a Bohr radius of about 1 nm~\cite{Qiu2013}, are expected to shift to significantly higher energies in NRs in view of quantum confinement effects, while the  E$_1$  and E$_2$ peaks appear at lower energies with respect to the bulk excitons. 
\begin{figure}
\begin{center}
\includegraphics[width=0.48\textwidth]{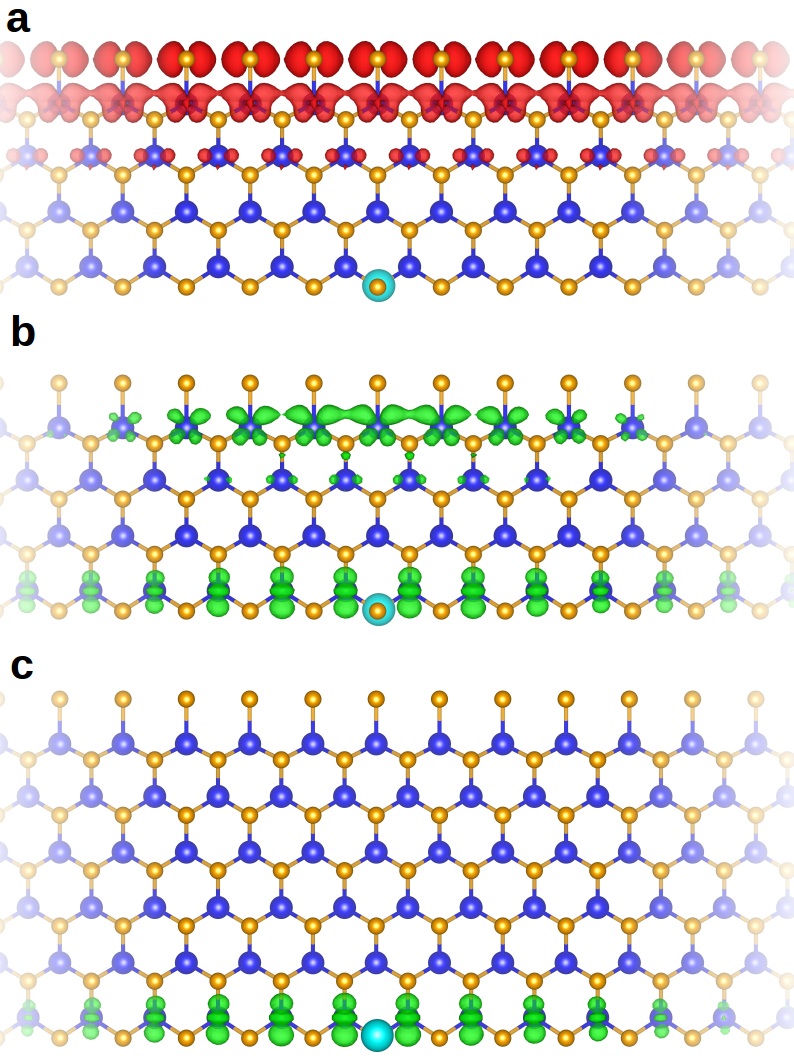}
\caption{
Square modulus of the exciton  wavefunctions for the strongest excitations contributing to the absorption peaks. For the $N=2$ ZZ-NR optical spectrum we show the results for both the E$_1$ (a) and E$_2$ (b) peaks while for the the $N=3$ case we show the E$_2$ (c) peak only, the wave-function of the E$_1$ peak remaining unchanged.
In all cases the excitonic wavefunction represents the electron distribution (red (green) for the E$_1$ (E$_2$) peak) for a given hole position (cyan circle) that has been fixed in order to maximize the density of the initial states contributing to the main transitions building the corresponding exciton state (see red and green full-circles in Fig.~\ref{bands_and_absorption_spectra}a and c).
\label{wave-functions}}
\end{center}
\end{figure}
In order to gain insight into the nature of these excitations, we analyze the composition of the E$_1$ and E$_2$ peaks. The E$_1$ peak originates from a single excitation, which mainly arises from the combination of transitions between metallic edge states almost at the center of the Brillouin Zone (BZ), as shown by red full (empty) circles in Fig.~\ref{bands_and_absorption_spectra}a that highlight the valence (conduction) states  involved. In particular, occupied and empty states are localized on opposite edges, thus forming an ``inter-edge'' exciton, with the valence states localized on the S-edge and the conduction states on the S-dimer terminating the Mo-edge (see the excitonic  wave-function in Fig.~\ref{wave-functions}a and the electronic-density distributions of the states involved in the single-particle transitions as reported in the SM~\cite{SM}). 
By enlarging the width of the nanoribbon ($N=3$) the single-particle states contributing to the formation of the E$_1$ peak will still be the same (localized on opposite edges), leaving a largely unchanged  exciton wave-function analogous to the one shown in Fig.~\ref{wave-functions}a for the $N=2$ case. However, the vanishingly small overlap results in a suppression of the corresponding absorption peak (see inset of Fig.~\ref{bands_and_absorption_spectra}d).

A similar analysis shows that the E$_2$ peak is built up by several excitations, as indicated by orange bars in Fig.~\ref{bands_and_absorption_spectra}b-d. By analyzing the strongest contribution to the peak, we recognize an ``intra-edge'' character of the exciton, which mainly involves transitions located at about one fifth of the BZ, corresponding to edge-states localized on the Mo-edge, and transitions near the edge of the BZ, coupling an extended valence state and a conduction state located on the S-edge (see green full (empty) circles in Fig.~\ref{bands_and_absorption_spectra}a-c, the exciton  wave-functions in Fig.~\ref{wave-functions}b-c and the electronic-density distributions of the states involved in the single-particle transitions as reported in the SM~\cite{SM}). The electronic density of the exciton wave-function present on the Mo-edge for the E$_2$ in the case of $N=2$ (Fig.~\ref{wave-functions}b) vanishes by enlarging the width of the nanoribbon ($N=3$) resulting in an excitation localized on a single edge (Fig.~\ref{wave-functions}c).

The above analysis allows us also to rationalize the optical properties when varying the NR width. Indeed, by comparing the spectra for $N=2$ and $N=3$ (Fig.~\ref{bands_and_absorption_spectra}b and d) it is clear that  the oscillator strength of E$_1$ becomes vanishingly small for $N=3$ (see inset), while E$_2$ remains optically active, with the contribution to the peak being limited to a single excitation for the wider NR. These results can be traced back to the fact that E$_1$ presents  spatial distributions for electrons and holes that are localized on opposite edges, with a suppression of their overlap --and thus of the exciton optical activity-- for sufficiently large widths. On the contrary, E$_2$ retains  a finite oscillator strength because the electron and hole spatial distributions are localized on the same side. For the same reason, the analysis of the composition of E$_2$ shows that also the energy position, binding energy and character of the excitations remain nearly the same by increasing the width.

We remark that, according to previous investigations~\cite{Bollinger2003,gibertini15},  mid-gap  edge states arise irrespective of the  details of the edge orientation (ZZ or AC) or  passivation. Since the main excitonic properties investigated for the S-dimer termination are intimately related to the presence of such  edge states, we expect them to be a common feature for NRs with different edge passivations.
To support this, we have computed the $GW$-BSE optical spectrum also for the S-monomer and the bare-edge terminations~\cite{SM}. In all cases, both inter- and intra-edge excitons have been identified below the bulk onset.
As summarized in  Table~\ref{excitons_summary}, the energy positions and binding energies of E$_1$ and E$_2$ are almost independent of the Mo-edge termination (S-dimer, S-monomer and bare-edge).
For the intra-edge peak E$_2$, results for both $N=2$ and $N=3$ are reported, while the properties of the inter-edge peak E$_1$ are provided only for $N=2$ since it becomes optically dark when the NR width is further increased (see discussion above). 
Analogous calculations for AC NRs~\cite{Kim2015} show that similar edge-related excitons below the main bulk features are present also for different edge orientations, although with possibly slightly different energies.
We thus have that the E$_1$ and E$_2$ peaks exist independently of the edge termination or orientation, confirming the robustness and the universal character of these two excitations. In particular, the intra-edge E$_2$ exciton remains unaffected in larger system sizes and should thus be visible not only in NRs but also at the edge of larger samples, as a clear peak red-shifted with respect to the bulk A and B excitons.

In all cases considered here for MoS$_2$ ZZ-NRs, the edge states become metallic as a consequence of the polar discontinuity across the NR edge, irrespective of the  details of the edge passivation~\cite{gibertini15}.  The existence of excitons in such metallic systems~\cite{mahan67} is a very interesting feature, as the strong screening of Coulomb interactions in metals is supposed to suppress excitonic resonances on short time scales~\cite{cui14,miller14,pile14}.
Nonetheless, in reduced dimensionalities screening is less effective, leaving a residual electron-hole interaction to form bound pairs. For example, excitons were predicted~\cite{spataru04,deslippe07} and experimentally observed~\cite{wang07} in 1D semi-metallic carbon nanotubes. Also in MoS$_2$ ZZ-NRs, metallic screening from the polarization-induced edge states (correctly accounted for in our calculations even in the $q\to0$ limit as specified in \cite{SM}) is not effective in suppressing completely the edge-localized excitons identified here, although the binding energy is remarkably smaller than for bulk excitons.
\begin{table}[t]
\caption {Energies and binding energies (in eV) of the E$_1$ and E$_2$ excitonic peaks as a function of the Mo-edge termination and of the nanoribbon width $N$.
\label{excitons_summary}}
\vspace{2mm}
\begin{tabularx}{\linewidth}{l >{\centering\arraybackslash}X
c
>{\centering\arraybackslash}X >{\centering\arraybackslash}X }
\toprule
      &  $E_1\, (E_b)$ && \multicolumn{2}{c}{ $E_2\, (E_b)$ }\\ 
      \cmidrule{2-2}\cmidrule{4-5}
      & $ N=2 $      && $N=2$     & $N=3$\\ 
\midrule

Bare & 0.50 (0.23) &&  1.53 (0.36) & 1.47 (0.40)\\
S-monomer & 0.68 (0.25) && 1.51 (0.39) & 1.42 (0.42)\\ 
S-dimer & 0.55 (0.21) && 1.49 (0.41) & 1.49 (0.39)\\

\bottomrule
\end{tabularx}
\end {table}
In conclusion, we have studied the electronic and optical properties of zigzag MoS$_2$ NRs as a function of the NR width and of the Mo-edge termination by means of accurate $GW$-BSE computational methods.
We have found that these systems present distinct low-energy optical features that are strictly connected to the emergence of mid-gap states localized at the edges: an inter-edge excitonic peak, peculiar of ultranarrow ribbons, emerging from the interaction between states localized at opposite edges, and a width-independent, intra-edge excitation involving states localized on the same edge of the ribbon. 
The existence of the latter excitons is robust and universal, irrespective of the edge orientation and termination.  In particular, the width-indepedent intra-edge exciton is expected in the optical spectrum not only of narrow ribbons, but also of larger samples, as a clear, edge-localized peak at energies below the bulk 2D excitons. Remarkably, the similarities in the chemical properties and electronic band structure between all group-VI TMDs, such as MoSe$_2$ and WS$_2$, hint at these optical features being even more general.
In the specific case of zigzag nanoribbons considered here, the edge states involved in the excitations become metallic as a consequence of the polar discontinuity across the edges~\cite{gibertini15}, thus forming 1D dimensional wires of free carriers. The finite binding energy reported for the edge-localized excitons thus shows that metallic screening is not effective in 1D, reinforcing similar conclusions on semimetallic carbon nanotubes~\cite{spataru04,deslippe07}. 

We gratefully acknowledge stimulating discussions with Margherita Marsili and Giovanni Borghi. This work was financially supported by the MaX -- MAterials at the eXascale -- Centre of Excellence, funded by the European Union programs H2020-EINFRA-2015-1 and H2020-INFRAEDI-2018-1 (Grant No. 676598 and 824143, respectively). M.G.\ acknowledges support from the Swiss National Science Foundation through the Ambizione program. Computational resources were provided through the PRACE projects 2016163879 and 2016163963 and the CSCS production project s825.

\bibliography{MoS2_ribbon_final}

\clearpage
\include{SI_final}
\end{document}

%% file: SI_final.tex
\renewcommand{\thepage}{S\arabic{page}}
\renewcommand{\thesection}{S\arabic{section}}
\renewcommand{\thetable}{S\arabic{table}}
\renewcommand{\thefigure}{S\arabic{figure}}



\onecolumngrid

\begin{center}
\textbf{\large Supplemental Material for \\
 ``Intrinsic edge excitons in two-dimensional MoS$_2$''}
\end{center}

\setcounter{equation}{0}
\setcounter{figure}{0}
\setcounter{table}{0}
\setcounter{page}{1}
\setcounter{section}{0}

\twocolumngrid

\section{Methods} \label{methods}
Simulations of the ground-state properties of MoS$_2$ zigzag nanoribbons (NRs) were performed using a first-principles  implementation of density-functional theory (DFT) based on plane waves and pseudopotentials, as available in the \QE{} distribution~\cite{espresso09,espresso17}. The local density approximation (LDA) within the Perdew Zunger parametrization~\cite{pz81} was adopted for the exchange-correlation potential, together with norm conserving pseudopotentials, with a 65-Ry energy cutoff on the wavefunctions. The supercell size was set to twice the width of the ribbons considered along the transversal direction and to 20 \AA{} in the direction perpendicular to the ribbon plane, in order to prevent interaction between periodic replicas. Moreover, the electrostatic potential in the direction perpendicular to the ribbon plane was corrected by using the Otani-Sugino approach~\cite{otani2006}. The (whole) Brillouin zone (BZ) was sampled by using a 1 $\times 12 \times 1$ $\mathbf{k}$-point grid, and atomic positions were fully relaxed until forces were smaller than 26 meV/\AA{} while the equilibrium lattice parameter along the NR was obtained by fitting the equation of state with a third-order polynomial.

The quasiparticle and optical absorption properties were subsequently computed within the framework of many-body perturbation theory~\cite{onid+02rmp}, according to the $GW$-BSE scheme. Calculations were performed by using the \yambo{} code~\cite{yambo,sangalli2019many}. Quasi-particle corrections to the Kohn-Sham eigenvalues were calculated within the $G_0W_0$ approximation for the self-energy operator, where the dynamic dielectric function was obtained within the plasmon-pole approximation~\cite{godb-need89prl}. The Coulomb potential was truncated by using a box-shaped cutoff to remove the long-range interaction between periodic images and simulate isolated systems~\cite{rozz+06prb}. The BZ was sampled with a $1 \times 48 \times 1$ $\mathbf{k}$-point grid. A kinetic energy cutoff of 7 Ry was used to represent the screening matrix, and the sum-over-states in the calculation of the polarization function (Green's function) was truncated at 1000 (1000) bands, corresponding to 30 eV above the Fermi level.

The optical absorption spectra were then computed as the imaginary part of the macroscopic dielectric function $\varepsilon$ starting from the solution of the BS equation in order to take into account electron-hole interaction. The static screening in the direct term was calculated within the random-phase approximation, with the same parameters as for $GW$ calculations. 
Converged excitation energies were obtained considering respectively five valence and five conduction bands in the BSE matrix, together with the application of the Tamm-Dancoff approximation, after having verified that the correction introduced by coupling the resonant and antiresonant part was negligible. 

The full set of calculations presented in the main text was performed by neglecting spin-orbit (SO) coupling, which is expected not to alter the qualitative description of optical spectra, as shown in Section~\ref{SOI}. 

\section{Computational details}

For the $G_0W_0$ calculations, we set a tolerance of 10 meV in representative energy gaps at zone boundaries to choose the main parameters (number of bands for the sum-over-states and cut-off energy for the response function, see text) which have then been fixed to check the convergence of $G_0W_0$ band structures and BSE spectra with respect to the BZ sampling, as shown in Fig. S1. The $1 \times 48 \times 1$ $\mathbf{k}$-point grid mentioned in the previous section has been chosen as a good compromise in order to reduce the computational load~\footnote{Parallel calculations on the PIZ-DAINT machine at the CSCS supercomputing center have required up to 6000 nodes/hours for a typical $GW$-BSE single run.}; with such grid we find that the binding energy of the excitonic peaks differ by approximately 10 meV from the extrapolated value for the infinite grid limit.
\begin{center}
\begin{figure}[!htb]
\includegraphics[width=0.48\textwidth]{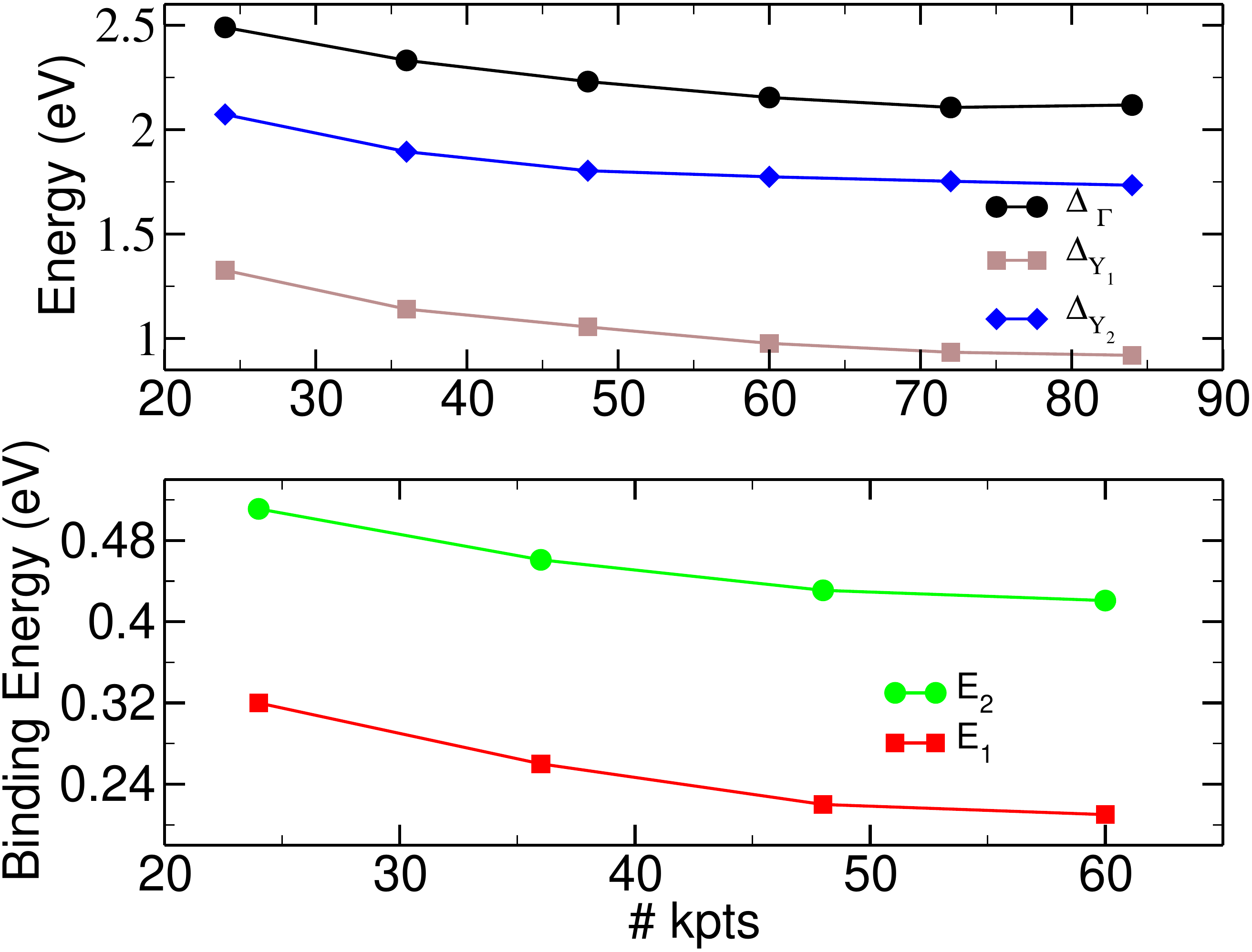}
\caption{Convergence analysis of the $GW$-BSE calculations {\it vs} $\mathbf{k}$-point grid. (Top) Energy gaps at zone boundaries as a function of the number of k-points along the ribbon axis for the case of $N=2$. The energy gaps $\Delta$ are calculated at $\Gamma$ and Y points, where the subscript Y$_1$ (Y$_2$) refers to the gaps at the Y point between the valence band maximum and the first (second) state above the Fermi level. (Bottom) Binding energy of the E$_1$ and E$_2$ excitons as a function of the number of $\mathbf{k}$-points along the ribbon axis for the case of $N=2$.
}
\label{convergence.picture}
\end{figure}
\end{center}
We remark that we need to compute self-energy corrections for the whole BZ, and for each band entering the BSE calculations, since approximated correction schemes, such as `scissor-and-stretching', cannot be applied in this case, as demonstrated by the plot of LDA vs QP energy values for the $N=2$ ribbon in Fig. S2(left panel). Interestingly, the metallic bands around the Fermi level show remarkably different corrections with respect to the bulk extended states, as can be seen from the comparison with the corresponding electronic structure in the right panel of Fig. S2.
\begin{center}
\begin{figure}[!htb]
\includegraphics[width=0.48\textwidth]{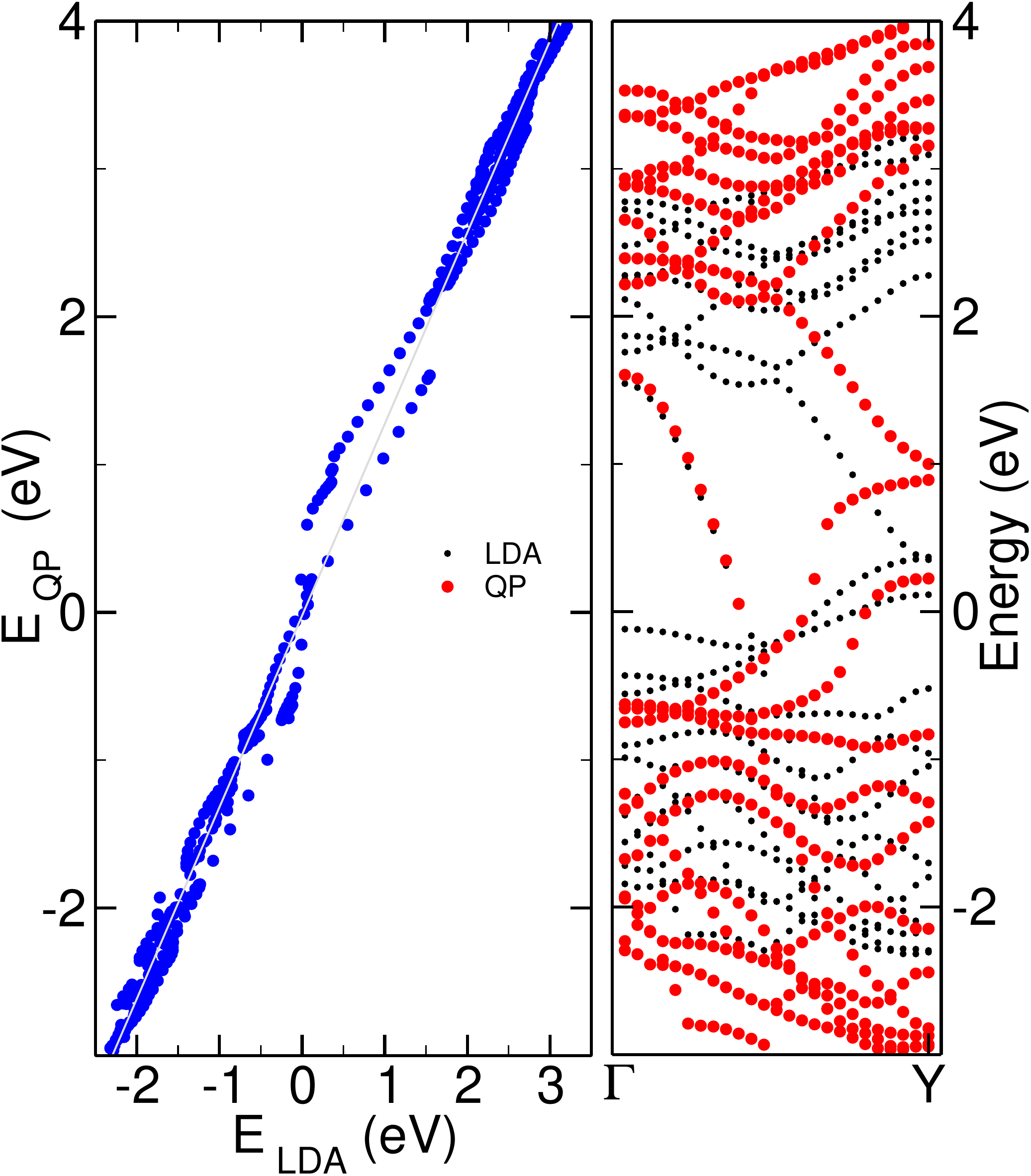}
\caption{(Left) Analysis of the LDA and quasiparticle (QP) enegies for the $N=2$ NR, where the grey line is a linear fit. The corresponding full electronic structures are reported in the right panel.}
\label{energy_analisys}
\end{figure}
\end{center}
%

\section{Inclusion of spin-orbit interaction} \label{SOI}
In order to check the validity of our results with respect to the inclusion of spin-orbit (SO) effects, we have computed the LDA electronic structure and the corresponding independent-particle (IP) optical spectrum in presence of SO interaction for the $N=2$ NR.
The comparison of the LDA electronic structure (Fig. S3, left panel) and the corresponding IP absorption spectrum with and without SO (Fig. S3, right panel) shows that the effect of SO coupling affects the spectrum only slightly, without changing its characteristic features. In particular, the two peaks investigated in the present work are still present, with a redshift of about 0.1 eV.

\begin{figure}[!tb]
\includegraphics[width=0.48\textwidth]{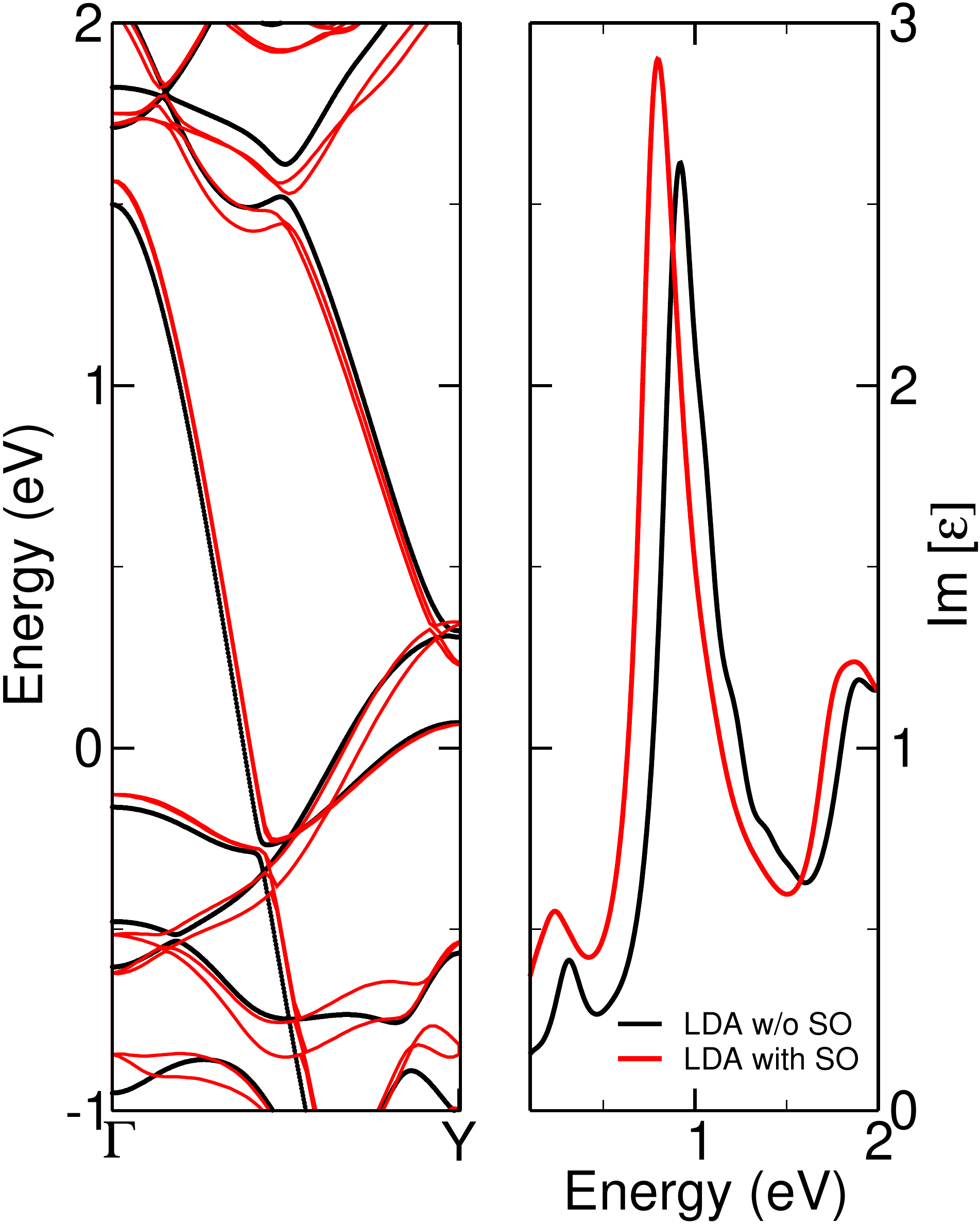}
\caption{Electronic band structure (left panel) and absorption spectrum (right panel) of the $N=2$ NR with (red curves) and without (black curves) SO coupling.}
\label{LDA.with.SO}
\end{figure}
%

\section{Treatment of the metallic screening} \label{Yukawa}
In order to include the metallic limit ($q \rightarrow 0$) in the description of the screened interaction for the construction of the BSE kernel, we have adopted a specific procedure that we discuss in the following.
The usage of a truncated potential to properly avoid spurious interactions among replicas implies that our system is represented as a quasi-1D system.
In the long-wavelength limit, the screened Coulomb interaction in 1D electronic systems can be expressed in terms of Bessel functions through the equation 
\begin{equation} \label{1D_Bessel}
    W(q)=\frac{4}{(q+q_0)^2R^2}\left[1-(q+q_0) R K_1((q+q_0)R)\right],
\end{equation} 
or alternatively as
\begin{equation} \label{1D_EI}
    W(q)=-\exp\left({(q+q_0)^2R^2}\right)Ei\left(-(q+q_0)^2R^2\right),
\end{equation} 
where $q_0$ and $R$ are two parameters (namely the inverse screening length and the typical transverse size of the system) while $K_1$ and $Ei$ are the modified Bessel function of second kind (with order 1) and the exponential integral function (see for example \cite{giuliani-vignale} for reference).
In our computational approach to build the BSE~\cite{yambo} the only point that is not well represented is the $q=0$ one (or the limit $q \rightarrow 0$) since for every finite $q$ the intraband transitions are taken into account fully ab-initio while for $q=0$ only vertical inter-band transitions are included.
In order to correctly describe the $q \rightarrow 0$ limit the $W(q=0)$ contribution is evaluated by fitting the numerical data of W (excluding the q=0 point) by Eq.~(\ref{1D_Bessel}). The result of the fitting is shown in Fig. S4.
Moreover, the fact that the screened interaction can be well represented close to $q=0$ with the 1D analytical expression confirms the quasi one-dimensional nature of our system.
\begin{figure}[!tb]
\includegraphics[width=0.48\textwidth]{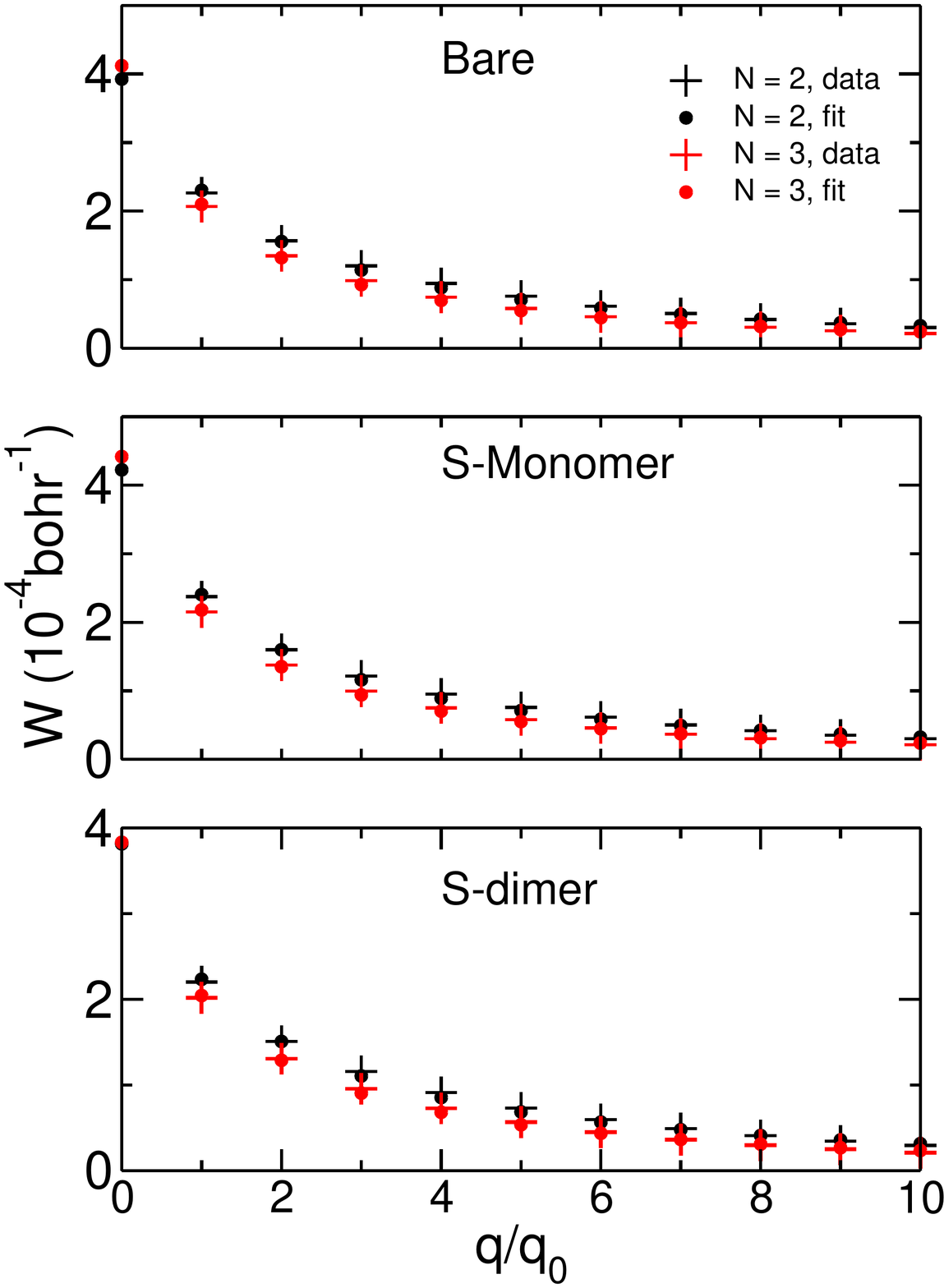}
\caption{Calculated (plus symbols) and fitted (full circles) values of the screened Coulomb interaction ($W$) as a function of the transferred momentum $q$ (expressed in units of $q_0$, the smallest value resulting from the discretization of the BZ). Starting from the top of the figure, we show results for the bare, S-monomer and S-dimer terminated Mo edges where black (red) curves refer to $N=2 (3)$. The fits have been performed according to Eq.~(\ref{1D_Bessel}) for data starting from $q=1$ and $W(q=0)$ has been evaluated accordingly. We can see that in all the cases the fitted values agree very well with the calculated ones.}
\label{W1D_fits}
\end{figure}
%
\section{Bare-edge, S-monomer and S-dimer terminations}
Before discussing the bare and the S-monomer termination we report the electronic density for the main single-particle transition contributing to the formation of the E$_1$ and E$_2$ peaks discussed in the main text in order to complement the information given by the square-modulus of the excitonic wave-functions. In Fig. S5 we show the side-view electronic density for valence ($\varphi_{vk}$) and conduction ($\varphi_{ck}$) bands building up the strongest excitonic states involved in the E$_1$ and E$_2$ peaks of the $N=2$ NR with S-dimer termination. We can clearly see that while the transition for the E$_1$ peak has an intra-edge character connecting left and right edges, the E$_2$ peak contains transitions that have an intra-edge nature connecting valence and conduction states located on the same edge of the ribbon.
\begin{figure}
\begin{center}
\includegraphics[width=0.48\textwidth]{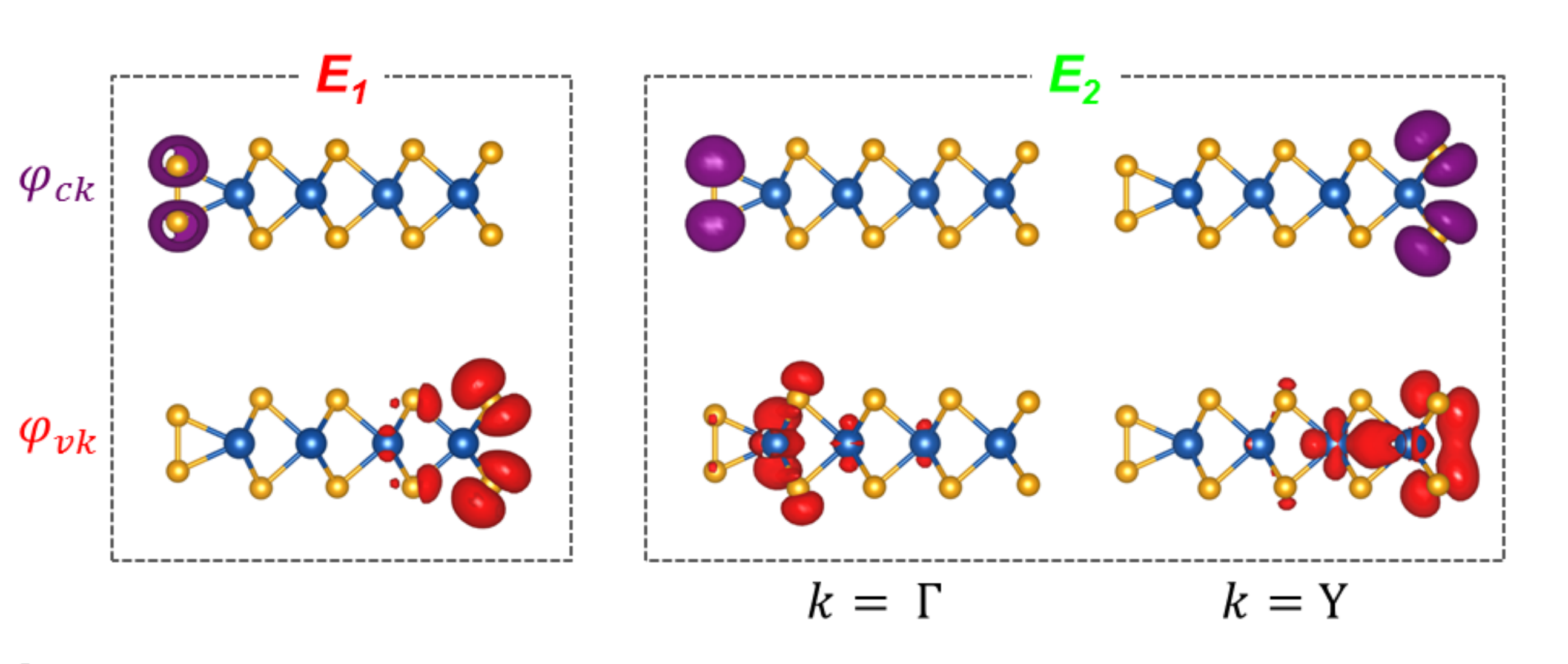}
\caption{Side-view of the electronic density of the single-particle states contributing to transitions building the E$_1$ (left) and E$_2$ (right) excitons for the $N=2$ NR with S-dimer termination. The densities corresponding to the valence (conduction) states are depicted in red (purple). For the E$_2$ exciton right and left-edge localized transitions are observed for points close to $\Gamma$ and Y in the reciprocal space.
\label{SP_WFs}}
\end{center}
\end{figure}

To address the edge independence of the investigated excitonic edge effects, we have performed a similar analysis to the one presented in the main text for S-dimer termination. In particular we have studied the behaviour of the optoelectronic properties for S-monomer and bare-edge terminations of the Mo-edge of the NRs.
In Fig. S6, S7 and Fig. S8, S9 we report the results for the bare and S-monomer termination respectively. They are presented as in the main text, where we have analyzed the results in terms of edge functionalization showing that the properties of the E$_1$ and E$_2$ excitons are almost insensitive to the specific edge termination.
We note that the optical spectrum of the $N=3$ NR with S-monomer termination has an additional excitonic peak E$_3$ close to the strongest one E$_2$, as indicated in purple in Fig. S8(d). Its composition in terms of single-particle transitions is indicated in Fig. S8(c) with the same color coding.
\begin{figure}
\begin{center}
\includegraphics[width=0.48\textwidth]{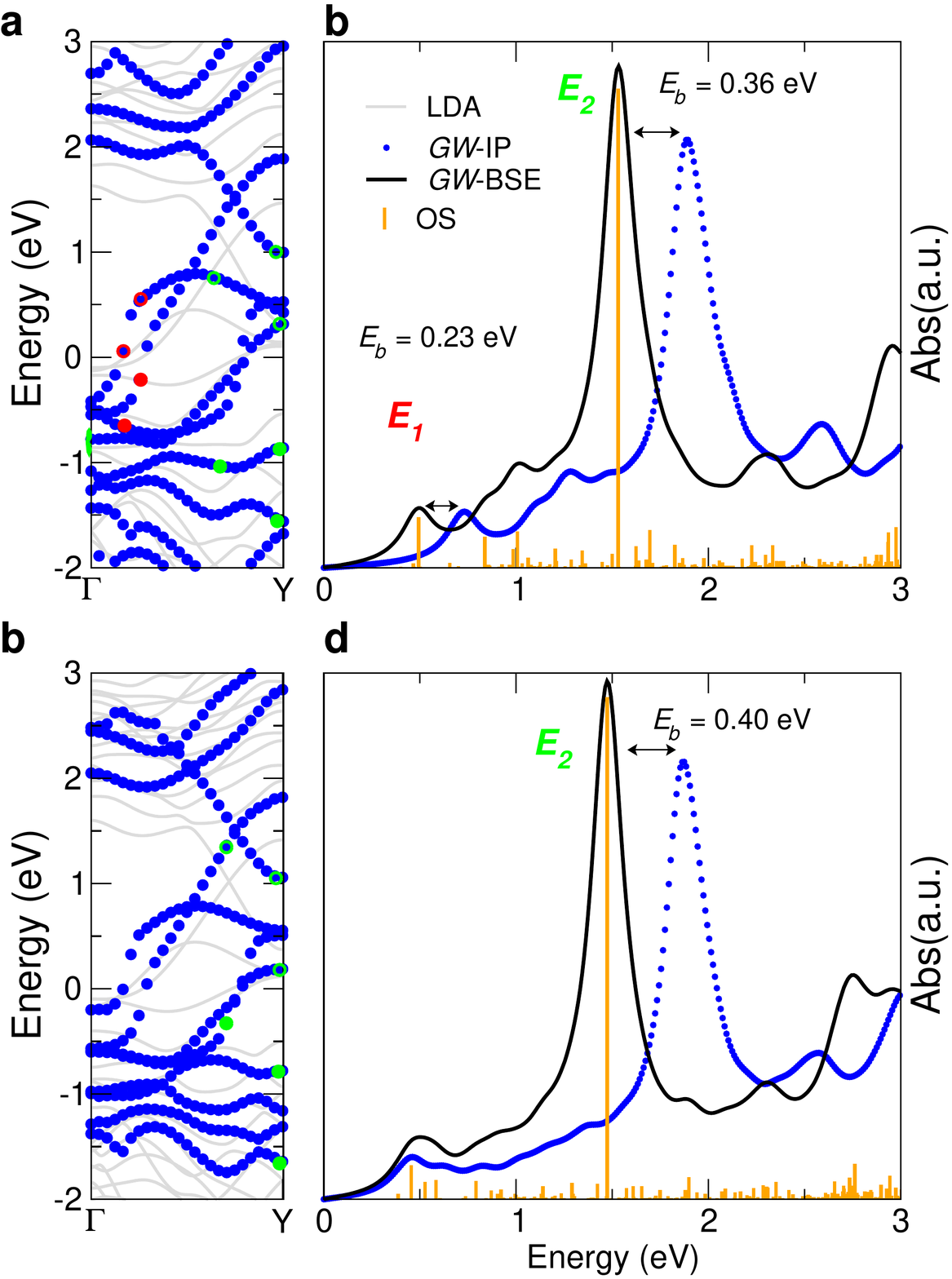}
\caption{Electronic band structures (a,c) and optical absorption spectra (b,d) in the case of bare Mo-edge termination for the $N=2$ (a,b) and $N=3$ (c,d) Z-MoS$_2$-NRs, respectively.
Panels a,c: Gray solid lines and blue dots correspond to the LDA Kohn-Sham and to the $G_0W_0$ bands, respectively, where the Fermi level has been set to zero.
Panels b, d: Blue-dotted curves ($GW$-IP) represent the optical absorption spectra as obtained according to the independent particle approximation by considering the $GW$ corrected energies and the LDA Kohn-Sham orbitals, whereas black solid curves ($GW$-BSE) correspond to the optical absorption spectra with the inclusion of electron-hole interaction within the $GW$-BSE scheme. In both cases, an artificial Lorentzian broadening of 0.1 eV has been used for convenience, while the relative oscillator strength of individual excitonic states is indicated by orange bars. The binding energies $E_b$ of the most relevant peaks, i.e. E$_1$ and E$_2$, are indicated in panels (b, d), while the peak composition in terms of single-particle transitions is indicated by red (E$_1$) and green (E$_2$) dots in panels (a,c), where full (empty) dot stands for occupied (empty) states.
\label{bands_and_absorption_spectra_bare}}
\end{center}
\end{figure}
\begin{figure}
\begin{center}
\includegraphics[width=0.48\textwidth]{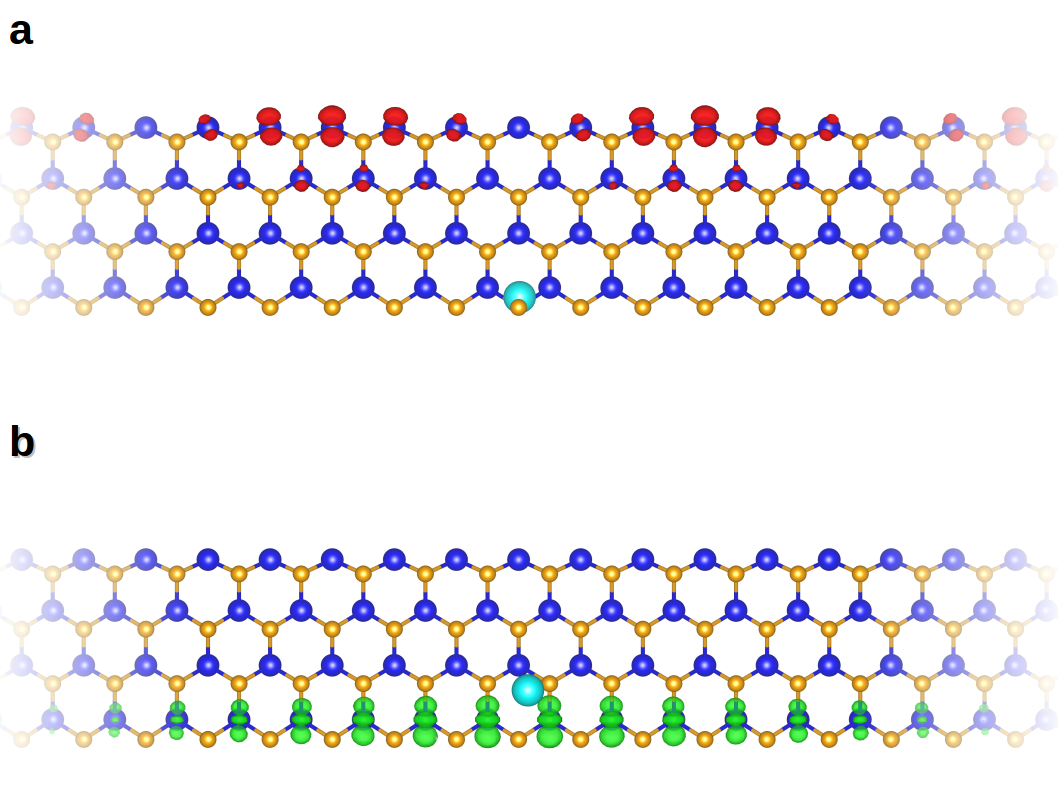}
\caption{Exciton square wavefunctions for the strongest excitations contributing to the E$_1$ (a) and E$_2$ (b) peak of the $N=2$ NR optical spectrum in the case of bare Mo-edge termination. The two excitonic square wavefunctions represent the electron distribution for a given hole position (cyan circle) that has been fixed in order to maximize the density of the initial states contributing to the main transitions building the corresponding exciton state (see red and green full-circles in Fig. S6(a,c)).
\label{wave-functions_bare}}
\end{center}
\end{figure}
\begin{figure}
\begin{center}
\includegraphics[width=0.48\textwidth]{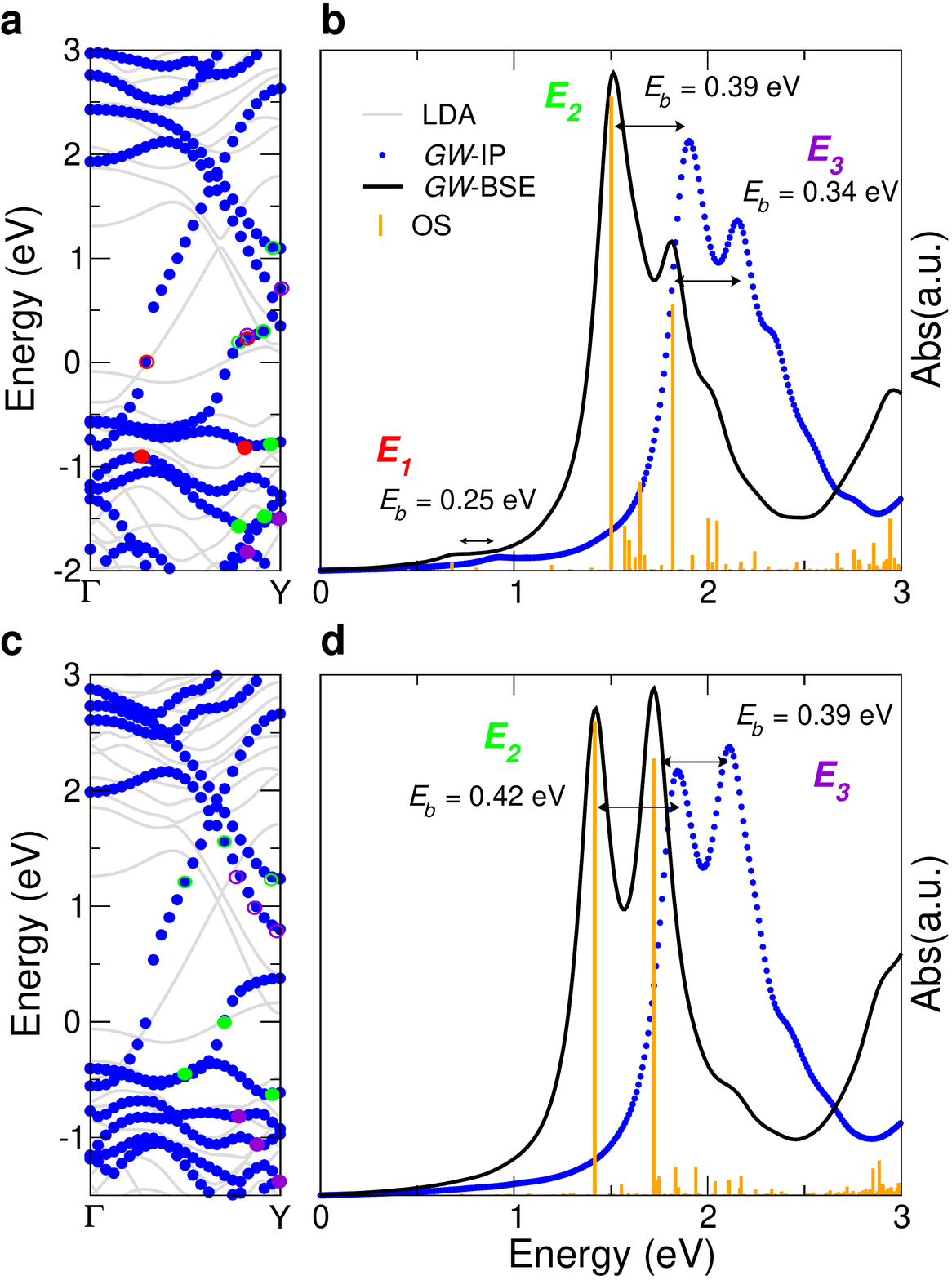}
\caption{Electronic band structures (a,c) and optical absorption spectra (b,d) in the case of S-monomer Mo-edge termination for the $N=2$ (a,b) and $N=3$ (c,d) Z-MoS$_2$-NRs, respectively.
Panels a,c: Gray solid lines and blue dots correspond to the LDA Kohn-Sham and to the $G_0W_0$ bands, respectively, where the Fermi level has been set to zero.
Panels b, d: Blue-dotted curves ($GW$-IP) represent the optical absorption spectra as obtained according to the independent particle approximation by considering the $GW$ corrected energies and the LDA Kohn-Sham orbitals, whereas black solid curves ($GW$-BSE) correspond to the optical absorption spectra with the inclusion of electron-hole interaction within the $GW$-BSE scheme. In both cases, an artificial Lorentzian broadening of 0.1 eV has been used for convenience, while the relative oscillator strength of individual excitonic states is indicated by orange bars. The binding energies $E_b$ of the most relevant peaks, i.e. E$_1$ and E$_2$, are indicated in panels (b, d), while the peak composition in terms of single-particle transitions is indicated by red (E$_1$) and green (E$_2$) dots in panels (a,c), where full (empty) dot stands for occupied (empy) states. 
\label{bands_and_absorption_spectra_monomer}}
\end{center}
\end{figure}
\begin{figure}
\begin{center}
\includegraphics[width=0.48\textwidth]{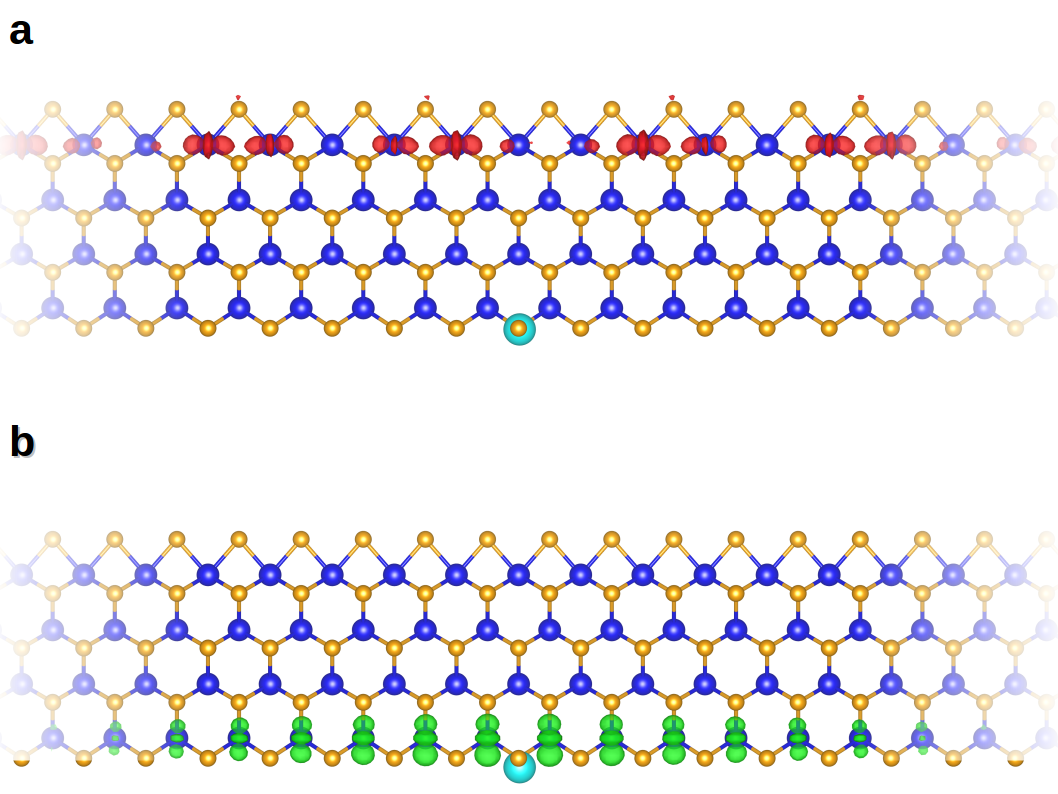}
\caption{Exciton square wavefunctions for the strongest excitations contributing to the E$_1$ (a) and E$_2$ (b) peak of the $N=2$ NR optical spectrum in the case of S-monomer Mo-edge termination. The two excitonic square wavefunctions represent the electron distribution for a given hole position (cyan circle) that has been fixed in order to maximize the density of the initial states contributing to the main transitions building the corresponding exciton state (see red and green full-circles in Fig. S8(a,c)).
\label{wave-functions_monomer}}
\end{center}
\end{figure}